\documentclass[letter, 12pt]{article}

\usepackage[margin=1in]{geometry}
\usepackage[numbers,sort&compress]{natbib} 
\usepackage{authblk} 
\usepackage{color}
\usepackage{amsmath}
\usepackage{amssymb} 
\usepackage{graphicx} 
\usepackage{fancyhdr} 
\usepackage{amsfonts}
\usepackage{tabularx} 
\usepackage[
singlelinecheck=false 
]{caption}

\def\be{\begin{equation}}
\def\ee{\end{equation}}
\def\bea{\begin{eqnarray}}
\def\eea{\end{eqnarray}}
\def\nn{\nonumber}

\def\ba{\begin{array}}
\def\ea{\end{array}}

\begin{document}

\renewcommand\headrule{} 

\title{\huge \bf{Axion star collisions with Neutron stars and Fast Radio Bursts}}
\author{Stuart Raby}
\affil{\emph{Department of Physics}\\\emph{The Ohio State University}\\\emph{191 W.~Woodruff Ave, Columbus, OH 43210, USA}}

\maketitle
\thispagestyle{fancy}
\pagenumbering{gobble} 

\begin{abstract}\normalsize\parindent 0pt\parskip 5pt
Axions may make a significant contribution to the dark matter of the universe.  It has been suggested
that these dark matter axions may condense into localized clumps, called ``axion stars."  In this
paper we argue that collisions of dilute axion stars with neutron stars, of the type known as ``magnetars," may be the origin of
most of the observed fast radio bursts.  This idea is a variation of an idea originally proposed
by Iwazaki.  However, instead of the surface effect of Iwazaki,  we propose a perhaps stronger
volume effect caused by the induced time dependent electric dipole moment of neutrons.
\end{abstract}

\pagenumbering{arabic} 

\newpage
\section*{Introduction}
Axions are the most natural solution to the strong CP problem \cite{Peccei:1977hh,Weinberg:1977ma,Wilczek:1977pj}.
There are significant phenomenological constraints on axions,  such that only so-called ``invisible axions" remain
as viable candidates (for a review, see \cite{Kim:2008hd}). These are known as the KSVZ axion \cite{Kim:1979if,Shifman:1979if} or
DFSZ axion \cite{Dine:1981rt,Zhitnitsky:1980tq}.  The supersymmetric version of invisible axions was given in \cite{Nilles:1981py,Kim:1983dt}.
Axions, with a decay constant $f_a \sim 10^{12}$ GeV can make up the dominant form of dark matter in the universe.  They form a smooth coherent
background state.   Laboratory experiments such as ADMX \cite{Rosenberg:2010zz,Carosi:2012zz} and light shining through walls \cite{Redondo:2010dp,Betz:2013dza} have been designed to look for them.  These are based
on the fact that an axion in a strong magnetic field produces light.   In a recent paper, Graham and Rajendran \cite{Graham:2013gfa}, have used the generic phenomenon that axions are responsible for a time dependent electric dipole moment of nuclei (see, for example, \cite{Pospelov:1999ha}).  They propose searching for this tiny effect.

On another line of reasoning,  it has been known that a self-gravitating Bose gas can form stable clumps of matter \cite{Kaup:1968zz,Ruffini:1969qy,Breit:1983nr,Tkachev:1991ka,Wang:2001wq}.  In the case of axions these clumps are termed ``axion stars"  \cite{Barranco:2010ib,Eby:2014fya,Chavanis:2011zm,Guth:2014hsa}.  These objects are very dilute with $M_* \sim 10^{-14} M_\odot$ and radius $R_a \sim 10^{-4} R_\odot$.  Recently the possibility of dense axion stars have been discussed \cite{Braaten:2015eeu}.  Note, if primordial density perturbations in the axion field leads to most of the axion background contained in axion stars then it will be very difficult to find axions in laboratory experiments.

In the case of axion stars, other methods for discovery are necessary.  One possibility was suggested by Iwazaki \cite{Iwazaki:2014wta}.  He uses   the electromagnetic coupling of axions, given by ${\cal L} \supset  \frac{\alpha}{f_a \pi} a(\vec{x},t) \vec{E} \cdot \vec{B}$  where $\vec{E}, \ \vec{B}$ are the electric and magnetic fields.   Neutron stars have strong magnetic fields and free electrons moving in the thin atmosphere on the surface. When an axion star enters the atmosphere of a neutron star, a time dependent electric field is produced by the axion star in the background magnetic field of the neutron star.  Then an immense amount of energy is released due to the time dependent coherent dipole oscillations
of the electrons.   The energy released is in the radio frequency (note $m_a c^2 = 10^{-5}$ eV corresponds to a frequency $\frac{m_a c^2}{2 \pi} = 2.5 $  GHz).  This suggests that the encounter of axion stars with neutron stars might be the source for Fast Radio Bursts [FRBs]  \cite{Lorimer:2007qn,Keane:2009cz,Thornton:2013iua,Spitler:2014fla}.  For a review on FRBs, see \cite{Katz:2016dti}.  There are only 17 known FRBs in the list.  Recently a repeating FRB has been discovered \cite{Spitler:2016dmz}.  This may make the chance encounter of an axion star with a neutron star an unlikely source for this phenomena,  but it does not exclude this possibility.

In this paper we consider the effect of the time dependent electric dipole moment for neutrons in the upper core of the neutron star.
These also radiate in the presence of an axion star and we argue that a significant portion of this radiation is coherent.   As a consequence, much more energy is available which can be released as FRBs.

 \section*{Axion stars}
Let us first note some of the parameters for, so-called, dilute axion stars.  They have a minimum radius, $R_a/R_\odot \sim 3 \times 10^{-4}$, with a maximum mass,  $M_a/M_\odot \sim 6 \times 10^{-14 \mp 4}$ for an axion mass, $m_a c^2 \sim 10^{-4 \pm 2}$ eV \cite{Chavanis:2011zm}.\footnote{$M_\odot \approx 2 \times 10^{30} Kg = 1.2 \times 10^{57} \ m_N$ and $R_\odot \approx 7 \times 10^8 m$.}
For later use we note that in a dilute axion star the central value of the axion field, $a_0$, for an axion star with maximum mass is given by $a_0/f_a \sim 10^{-7 \mp 2}$.  In general, we have $a_0/f_a \sim 10^{-7 \mp 2} (M_a/M_a^{max})^4$  where $R_a/R_\odot \sim 3 \times 10^{-4}  (M_a^{max}/M_a)$.

For a dense axion star, on the other hand, the minimum mass is approximately equal to $10^{-20 \mp 6} \ M_\odot$ with a minimum radius of order $R_a/R_\odot \sim 10^{-11 \mp 2}$.  It has a maximum mass of order $M_a/M_\odot \sim 1$ and maximum radius, $R_a/R_\odot \sim 10^{-5}$ \cite{Braaten:2015eeu}.   In this case we have $a_0/f_a \sim 1$.

The flux, $F_a$, of axion stars in the galaxy is roughly given in terms of the energy density of dark matter in the galaxy, $\rho_{DM} \sim 0.3 {\rm GeV/cm^3}$, times the virial velocity, $v_{vir}\sim 300 {\rm km/s}$, of objects in the galaxy.  Thus $F_a \sim \left( \rho_{DM}/M_a \right) v_{vir}$. For the calculations in this letter we take $m_a c^2 = 10^{-5}$ eV, corresponding to $f_a \sim 10^{12}$ GeV, which is in the regime required for a natural abundance of axion cold dark matter \cite{Kim:2008hd}.  This gives a dilute axion star with maximum mass, $M_a/M_\odot = 6 \times 10^{-12}$  or $M_a = 6.6 \times 10^{45}$ GeV/c$^2$.  Note, the more massive the axion star, the smaller the flux of axion stars.  In addition, the denser the axion star,  the smaller is its radius and thus the collision rate for axion - neutron star collisions is suppressed.   For both these reasons, we only consider dilute axion stars in the following.  In order to evaluate the rate of axion-neutron star collisions we need to know the cross-section for this process.  This cross-section is given by the area, $A = \pi b^2$,  where $b$ is the impact parameter for the collision.   Due to the classical counterpart of the Sommerfeld enhancement we have $b^2 = (R_{ns} + R_a)^2  \left[1 + \frac{2 G_N M_{ns}}{(R_{ns} + R_a) v_{vir}^2} \right]$, where $R_{ns} = 10$ km is the typical radius of a neutron star, $M_{ns} = 1.4 M_\odot$ is its mass and $G_N$ is Newton's constant.  The enhancement factor $\frac{2 G_N M_{ns}}{(R_{ns} + R_a) v_{vir}^2} \sim 20$ and we obtain $A = 3.2 \times 10^6  \left( \frac{R_{ns} + R_a}{220 {\rm km}} \right) $ km$^2$.  Assuming of order $10^9$ neutron stars in a galaxy we obtain an event rate per galaxy, \be R_{a-ns} \sim 10^{-6} (6 \times 10^{-12} M_\odot/M_a)\left( \frac{R_{ns} + R_a}{220 {\rm km}} \right) {\rm y}^{-1} . \ee Note, the radius of a dilute axion star increases as its mass decreases. Thus an axion star with mass, $M_a/M_\odot \sim 2 \times 10^{-13}$ and $R_a/R_\odot \sim 9 \times 10^{-3}$ would produce an event rate per galaxy of order, $R_{a-ns} \sim 10^{-3} y^{-1}$.  Now, following Iwazaki \cite{Iwazaki:2014wta}, we note that this is the observed rate for fast radio bursts [FRBs] \cite{Thornton:2013iua}.

\section*{Fast Radio Bursts and Axion stars}

  FRBs have been observed at a frequency, $\nu_{FRB} \sim 1.4$ GHz.  This happens to be about the axion oscillation frequencey, $\nu_a = \frac{m_a c^2}{2 \pi} = 2.5 \left( \frac{m_a c^2}{10^{-5} {\rm eV}} \right)$ GHz. Moreover, the observed frequency will be red shifted.   Iwazaki discussed the FRBs produced when the axion star interacts with the strong magnetic field in the thin atmosphere of the neutron star.  It causes coherent dipole oscillations of the electrons which then radiate in the frequency set by the axion star over a time scale of order a milli-second or less.   In the following we describe an additional effect and perhaps the dominant effect.  FRBs are most likely due to events at redshifts of 0.5 to 1.  They are associated with a huge release of energy of order $10^{38} - 10^{40}$ ergs or $6.25 \times (10^{40} - 10^{42})$ GeV corresponding to a rate of energy production, $P$, exceeding $6.25 \times (10^{43} - 10^{45})$ GeV/s.

\section*{Neutron stars}

The outer core of neutron stars contains mostly neutrons with 3 to 5\% free protons and electrons \cite{Potekhin:2011xe}.  The neutron star cools to a temperature, $T \sim 10^6 {^\circ}K$ with a strong magnetic field, $B \geq 10^{12}$ G.  When the axion star enters the outer core of the neutron star it induces a time dependent electric dipole moment in the neutrons, $d(t) = d_0 \sin(\omega t)$ with $\omega = m_a c^2/\hbar$.   We have $d_0 = 2.4 \times 10^{-16} \left(\frac{a_0}{f_a}\right)$ e cm \cite{Pospelov:1999ha,Graham:2013gfa} and we take $a_0/f_a =  10^{-10}$.  This time dependent electric dipole moment radiates radio waves with frequency $\nu_a$. The total power radiated is given by \be P = \frac{(m_a c^2)^4 \alpha}{3 \pi \hbar} \left(\frac{d_0 N_n \rho}{e \hbar c}\right)^2  \ee where $\alpha = 1/137$ is the fine structure constant, $N_n$ is the number of neutrons affected by the axion star and $\rho$ is the net polarization of the neutron spins given by $\rho = \frac{N_n \uparrow - N_n \downarrow}{N_n \uparrow + N_n \downarrow}$.\footnote{We expect the polarized neutrons to radiate coherently \cite{Dicke:1954zz}.}  The neutrons are polarized by the strong magnetic field.  The magnetic field inside a neutron star has been modeled \cite{Braithwaite:2005xi,Ciolfi:2014jha}.   It includes both poloidal and toroidal components.   Neutrons have a magnetic dipole moment, $|\mu_n| \approx 6 \times 10^{-14}$ MeV/T.  Thus $2 |\mu_n| B = 1.2 \times 10^{-2} (B/10^{15} {\rm G})$ MeV  and the Boltzmann factor determining the ratio of spin up to spin down neutrons, $e^x$ is given by $ x \equiv \frac{2 |\mu_n| B}{k T} = 1.4 \ (B/10^{15} {\rm G})/(T/ 10^8 {^\circ} {\rm K})$  with $k T = 8.6 \ (T/ 10^8 {^\circ}{\rm K})$ keV.\footnote{The neutrons in the neutron star are believed to be in a $^3P_2$ superfluid state \cite{Schwenk:2003bc,Page:2010aw}. Thus the neutron spins are alligned in the condensate.  In addition we have assumed a large magnetic field, associated with that of neutron stars, known as magnetars.}   The net polarization is then $\rho = \tanh(x/2) \approx 0.6$.  We then find the power radiated given by
\bea  P = & 1.8 \times 10^{55} \ \left( \frac{m_a c^2}{10^{-5} \ {\rm eV}} \right)^4 \left( \frac{N_n}{1.7 \times 10^{57}} \ \frac{\rho}{0.6} \right)^2  \ {\rm GeV}/s & \\
= & 2.9 \times 10^{52} \ \left( \frac{m_a c^2}{10^{-5} \ {\rm eV}} \right)^4 \left( \frac{N_n}{1.7 \times 10^{57}} \ \frac{\rho}{0.6} \right)^2 \ {\rm erg}/s. & \nn \eea  Since we have assumed that the total mass of the axion star is $M_a/M_\odot \sim 2 \times 10^{-13}$ or $M_a \sim 2.2 \times 10^{44}$ GeV/c$^2$, the star loses all its mass to radiation in a small fraction of a second.   This pulse of energy would typically last much less than a milli-second. Note, however, that the rate depends sensitively on the axion mass, the net polarization, $\rho$, of the neutrons in the neutron star core (which depends on the magnitude of the magnetic field and the temperature in the neutron star), the fraction of the neutron star covered by the axion star, i.e. $f = \frac{N_n}{1.7 \times 10^{57}}$
and the timescale for the traversal of the axion star through the neutron star.  The last two effects depend crucially on the the amount of stretching of the axion star due to tidal forces and the velocity of the axion star.  We will comment on the issue of tidal forces in the next section.

Of course, the next question is does this radiation escape from the neutron star.  This is a very difficult question which would require serious numerical simulations.  But in the absence of this analysis, let me make some simplifying assumptions.   There are two properties of the interior neutron star which can dramatically affect the result.  First there is a huge magnetic field which is most likely a combination of poloidal and toroidal. Secondly, about 1\% of the outer core of the neutron star is a plasma of ionized protons and electrons.  It is believed that the protons in the outer core of a $10^{12}$ G neutron star are in a superconducting state.  As a result, the magnetic fields either form a lattice of Abrikosov flux tubes for a type II superconductor or an intermediate state of normal and superconducting matter for a type I superconductor \cite{Huebener:1974ui,PhysRevB.57.3058}.  In either case,  electromagnetic waves would be severely constrained.   However, if the interior magnetic field is large enough, $B > 2 \times 10^{16}$ G, then it is believed that the core is not superconducting \cite{Sinha:2015bva,Sinha2015}.  Moreover, for smaller magnetic fields, $ 10^{15} \ {\rm G} \leq B \leq 2 \times 10^{16} \ {\rm G}$,  the core may be only partially superconducting \cite{Sinha:2015bva,Sinha2015}.  Hence radiation may then propagate in this non-superconducting core. Neutron stars with surface magnetic fields of order $10^{15}$ G are known as ``magnetars."   It is believed that magnetars may be just as numerous as neutron stars with magnetic fields in the $10^{12}$ G range \cite{Duncan:1992hi,Thompson:1993hn,Thompson:1995gw,Duncan:2003,Turolla:2015mwa}.

The magnetic field might make it easier for the radiation to escape.  It can be shown that
the optical opacity of a medium is severely suppressed in the presence of a strong magnetic field \cite{Canuto:1971cd}.  In the
fully ionized atmosphere of a neutron star the opacity has been evaluated in Ref. \cite{Potekhin:2002nk}.  The absorption cross-section for transverse electromagnetic waves with $\alpha = \pm 1$ and frequency $\omega$ is given by (Eqn. 51, Ref. \cite{Potekhin:2002nk})
\be \sigma_\alpha^{ff} \approx \frac{\omega^2}{(\omega + \alpha \omega_{ce})^2 (\omega - \alpha \omega_{cp})^2 + \omega^2 \tilde \nu_\alpha^2} \frac{4 \pi e^2 \nu_\alpha^{ff}}{m_e c} \ee  where $\omega_{ce}, \ \omega_{cp}$ are cyclotron frequencies for electrons and protons, respectively.   For $\omega \ll \omega_{ce}, \ \omega_{cp}$ the absorption cross-sections vanishes $\propto \omega^2$.   In Fig. 7, Ref. \cite{Potekhin:2002nk}, the opacity was calculated for a plasma with
density, $\rho = 500$ g/cm$^3$, temperature, $T = 5 \times 10^6 {^\circ}$ K and magnetic field, $B = 5 \times 10^{14}$ G.  These results will change with a stronger magnetic field.   At magnetic fields of order $10^{16}$ G, the proton cyclotron frequency is two orders of magnitude larger than the 3.15 keV of Fig. 7, Ref. \cite{Potekhin:2002nk}.  Moreover, we are interested in radio frequencies of order $10^{-5}$ eV, which are much lower.

Assuming we can scale the opacity of Fig. 7, Ref. \cite{Potekhin:2002nk} (defined for $T \sim 5\times 10^6 {^\circ}$K and $B \sim 5 \times 10^{14}$ G) from $\kappa_0 \sim 10^{-4.5} cm^2/g$ at $\hbar \omega_0 = 10^3$ eV to lower frequencies with $\kappa(\hbar \omega_a) = (\frac{\hbar \omega_a}{\hbar \omega_0})^2 \ \kappa_0$  (where $\hbar \omega_a = m_a c^2 = 10^{-5}$ eV), we find an extinction coefficient $\rho \kappa = \tau^{-1}$ with $\tau \sim 32 \ (\frac{\rho}{10^{14} \ g/cm^3})^{-1} \ km$  for a core density, $\rho \sim 10^{14}$ g/cm$^3$.\footnote{For a stronger magnetic field the frequency $\hbar \omega_0$ would increase. Also note, the core temperature is actually closer to $T \sim 10^8 {^\circ}$K.}   Clearly most of the energy emitted by that fraction of the axion star which covers the outer core, located at approximately 1 km below the surface of the neutron star, makes it out of the neutron star.  The rest of the energy goes into heating the plasma.
Finally, there is necessarily a back reaction on the axion star, as its mass is converted into radiation.

\section*{Tidal distortion of the axion star}

Once the axion star approaches a distance $R_{tidal}$ from the neutron star given by \be R_{tidal} \approx (\frac{M_{ns}}{M_a})^{1/3} \ R_a, \ee i.e, when the gravitational force at the surface of the axion star is balanced by the tidal forces due to the neutron star, then the axion star stretches in the direction of the neutron star. If it approaches the neutron star with zero impact parameter, $b$, the radio burst may be prolonged \cite{Pshirkov:2016bjr}, of order a second.  However, if it grazes the neutron star at an impact parameter, $b \gg R_{ns}$, then only a fraction of the axion star will actually cross the neutron star.  Then the crossing timescale can be of order milli-seconds.

\section*{Conclusion}

In this paper we discuss the possibility that FRBs are caused by the collisions of dilute axion stars with
neutron stars of the type known as magnetars.   Unlike the previous analysis of Iwazaki \cite{Iwazaki:2014wta} which focused on axion stars producing electric dipole radiation from the thin atmosphere of a neutron star, the proposed mechanism can be a much larger volume effect having to do with the
induced time dependent electric dipole moment of the neutrons interior to the star.   Further calculations are needed to evaluate the viability of the proposed mechanism.

\section*{Acknowledgments}

I acknowledge illuminating conversations with E. Braaten, C. Hirata, R. Furnstahl, T. Thompson and  H. Zhang.
I also received partial support for this work from DOE grant, DE-SC0011726.

\clearpage
\newpage

\bibliographystyle{utphys}
\bibliography{bibliography}

\end{document}